\documentclass[aps,floatfix,amsmath,nofootinbib,10pt,]{revtex4}
\usepackage{graphicx}
\newcommand{\be}{\begin{equation}}
\newcommand{\ee}{\end{equation}}
\newcommand{\bea}{\begin{eqnarray}}
\newcommand{\eea}{\end{eqnarray}}

\begin{document}

\title{Extra Dimensions, Dilaton and Dark Energy}
\author{Tongu\c{c} Rador}
\email[]{tonguc.rador@boun.edu.tr}
\affiliation{Bo\~{g}azi\c{c}i University Department of Physics, 34342 Bebek, \.{I}stanbul, Turkey\\}
\affiliation{\.{I}zmir Institute of Technology, Department of Physics, Urla 35430, \.{I}zmir, Turkey\\}

\date{\today}

\begin{abstract}
In view of the recent observations showing that the universe is accelerating we discuss dilaton and radion stabilization from a  phenomenological perspective  using perfect fluid sources. One general conclusion we present is that the pressure coefficient along extra dimensions should be $-2$ if that of the observed dimensions is $-1$, the latter mimicking a cosmological constant compatible with experimental data. The conditions on the dilaton coupling are similarly strong: we find that if the  coupling of the dilaton $\phi$ to  fields other than gravity is of the form $\sqrt{-g}\;e^{(a-2)\phi}\;\mathcal{L}$ where $\mathcal{L}$ representing all other fields yielding the mentioned fluid, $a$ must be $-2$ if space-time dimensionality is $10$. Within our approach these conditions result taking constant radion and dilaton at the level of the  equations of motion. To ameliorate this we also discuss how dynamical stabilization may be achieved with a simple variant in which a dilaton potential is added in the picture where the mentioned constraints are shown to remain.
\end{abstract} 

\maketitle

\section{Introduction}
Recent observations indicate that the expansion of the universe is accelerating and the data is compatible with  a cosmological constant, $\Lambda$, as the responsible actor. The value of $\Lambda$ turns out to be many orders of magnitude below the canonical estimate from quantum theoretical considerations. If one restricts the study to non-quantum approaches $\Lambda$ could be seen just as another fundamental constant of physics. Surely even the non-quantum cosmology is beset with fine tuning problems such as the cosmic coincidence of the onset of acceleration
\footnote{ {A name better suited for this "Why Now?" question, in view of its intriguing nature, is {\em cosmic scandal} as coined in \cite{carroll1}. We also refer to this article for arguably more functional names for {\em dark energy.}}}. 
Nevertheless this is a problem somewhat unrelated to the {\em ease} by which the cosmological constant accommodates the observations. Such could be the view of a pragmatic who ignores aesthetics.
 
However even the pragmatic would feel a somewhat pronounced unease in trying to reconcile acceleration of observed space with the assumption of extra dimensions  since the naive introduction of $\Lambda$ to the d-dimensional Einstein-Hilbert action mandates a time dependent compactification radius \footnote{ {In fact if a cosmological constant is the only actor then extra dimensions expand and accelerate in the same way as the large dimensions; an expected fact in view of the isotropy of a cosmological constant.}}  in clear disconcert with stringent bounds on the cosmological evolution of fundamental constants.  

An important ingredient of string inspired extra dimensional theories is string/brane gas cosmology\footnote{A possibly incomplete list of references for the literature is given in \cite{BV}-\cite{sgc36}.}. This framework is rather successful for cosmology of the very early universe and can even be a candidate to replace the inflationary paradigm in that it also solves the problems of standard cosmology  and yields the same type of spectrum for density  perturbations \cite{rev3,scale}.  To raise an intriguing point let us ignore acceleration for a moment and assume that the universe expands as if it is dominated by objects which exert no pressure along observed dimensions. In such a scenario of the late universe string/brane gas cosmology is as successful compared to its phenomenology for early times, if not aesthetically better. In fact it can not
only accommodate a static radius for extra dimensions
and constant dilaton \footnote{ {Dilaton stabilization is slightly more involved than that of the radion since for the latter T duality provides a natural mechanism. Nevertheless this can be achieved in a multitude of ways. Simplest one being a dilaton potential. On the other hand a single agent can also achieve both radion and dilaton stabilization by invoking the use of S duality as for instance done in \cite{sgc22.5} via $(m,n)$ strings. At any rate a lagrangian needs some sort of dilaton potential for stabilization.}} 
but also in doing so has a working idea to explain the number of observed dimensions
\footnote{The seed of this idea is of course related to the work of Brandenberger and Vafa \cite{BV}. Nevertheless it was shown in \cite{sgc25} that stabilization of radion and dilaton also fixes the dimensionality of the observed space assuming some dimensions got large {\em somehow}. The ideas that resulted in \cite{sgc25} were slowly taking shape in earlier works \cite{sgc19}-\cite{sgc24}.} and has for instance arguments on the possibility for dark matter of string/brane origin \cite{rev2}-\cite{dark3},\cite{sgc16}. However the acceleration of the observed dimensions poses a challenge  to string/brane gas cosmology in view of the fact its constituents generally behave like pressureless dust along observed dimensions at late times. Thus it is of crucial importance to find an element compatible with string/brane gas cosmology that would stabilize extra dimensions and the dilaton while allowing our observed universe to expand in an accelerating fashion, desirably commensurable with a cosmological constant dynamics in a four dimensional point of view\footnote{This means that the observed dimensions expand exponentially. While as things stand for now this is not a must it is possibly the simplest scenario invoking Occam's razor.}. Surely one can also contemplate extra dimensional theories not motivated by strings and these are not free of the mentioned challenge. In this paper we follow a phenomenological approach: we assume the desired cosmological evolutions of fields and from that we infer conditions on the parameters of a general dilatonic extra-dimensional theory enriched with dilaton coupling to fields other than gravity. 

\section{A Simple Observation}{\label{secII}}

Let us assume the following action which can also be motivated by the low energy limit of  string theory;

\be\label{eq1}
S=\int\;dx^{d}\sqrt{-g}\;e^{-2\phi}\left[R+4(\nabla\phi)^{2}+e^{a\phi}\mathcal{L}\right]\;.
\ee

\noindent Within a cosmological scenario the metric can be chosen to be,

\be
ds^{2}=-dt^{2}+e^{2B(t)}d{\mathcal{O}_{m}}^2+e^{2C(t)}d{\mathcal{E}_{p}}^{2}\;,
\ee

\noindent where $d{\mathcal{O}_{m}}^2$ and $d{\mathcal{E}_{p}}^{2}$ represent the line elements of the $m$ and $p$ dimensional observed and extra dimensional manifolds ${\mathcal{M_{O}}}$ and ${\mathcal{M_{E}}}$ respectively. The quantities $B(t)$ and $C(t)$ are the corresponding scale factors and we have $d=1+m+p$.  For string theory applications one would take $d=10$. Unless otherwise stated we will, in this work,  assume that ${\mathcal{M_{O}}}$ and ${\mathcal{M_{E}}}$ are flat. In such a cosmological approach one can also assume a perfect fluid form

\be{\label{eqeben}}
\mathcal{L}=-2\rho\;\;.
\ee

\noindent Here $-2\sqrt{-g}\rho$ is assumed to yield a conserved energy-momentum tensor via 
the usual construction 

\be
T_{\mu\nu}=-\frac{1}{\sqrt{-g}}\frac{\partial \sqrt{-g}\mathcal{L}}{\partial g^{\mu\nu}}\;,
\ee

\noindent and thus will yield

\be
\rho=\rho_{i}\exp\left[-(1+\omega)mB-(1+\nu)pC)\right]\;\;.
\ee

\noindent Where $\omega$ and $\nu$ are the pressure coefficients along ${\mathcal{M_{O}}}$ and ${\mathcal{M_{E}}}$ respectively. The equations of motion for this  model are therefore\footnote{ {The model given by lagrangian (\ref{eq1}) along with the choice (\ref{eqeben}) have also been studied the way it is presented here in \cite{sgc22}. Later it was generalized in \cite{sgc27}.}}
\begin{subequations}
\bea
\ddot{B}+\dot{k}\dot{B}&=&e^{a\phi}\left(\omega-\tau\right)\rho\;,\label{bieq}\\
\ddot{C}+\dot{k}\dot{C}&=&e^{a\phi}\left(\nu-\tau\right)\rho\;,\label{cieq}\\
\ddot{\phi}+\dot{k}\dot{\phi}&=&\frac{1}{2}e^{a\phi}\left[T-(d-2)\tau\right]\rho\;,\label{phieq}\\
\dot{k}^{2}&=&m\dot{B}^{2}+p\dot{C}^{2} + 2 e^{a\phi}\rho\;,\label{refk1} \\
\dot{k}&\equiv& m\dot{B}+p\dot{C}-2\dot{\phi}\;.\label{refk2}
\eea
\end{subequations}
\noindent where $T=-1+m\omega+p\nu$, the trace of the energy-momentum tensor divided by $\rho$ and $\tau=(a-2)/2$. One can show that the signature of $\dot{k}$ is a constant of motion for positive $\rho$ and for $\dot{k}<0$ the solutions will be singular in finite proper time as was also argued in \cite{sgc22} and thus one should confine the study to $\dot{k}>0$.

Clearly if we would like to have a constant dilaton and radion, the right hand sides of (\ref{cieq}) and (\ref{phieq}) must vanish. Now assuming the observed dimensions expand in an accelerated fashion as $B(t)=Ht+B_{o}$, mimicking a four-dimensional cosmological constant and hence yielding $\omega=-1$, it is a simple exercise to show that one must have
\begin{subequations}{\label{hemhem}}
\bea
\nu &=& -\frac{m+1}{m-1}\;,\\
a&=&-\frac{4}{m-1}\;.
\eea
\end{subequations}

In a pure Einstein gravity context, that $\nu=-2$ for $m=3$ was shown in \cite{eksi2ben} for flat ${\mathcal{M_{O}}}$ and ${\mathcal{M_{E}}}$. There it was also shown that dynamical stabilization of the radion could be established via a curvature term for ${\mathcal{M_E}}$ next to our $\rho$. A curvature term for extra dimensions is a plausible companion to $\rho$ since in a perfect fluid approach it represents a term with $\omega=-1$. In short what was shown  in \cite{eksi2ben}  was that ${\mathcal{M_{E}}}$ has to have negative curvature and $\nu\leq -2$ to have stabilization in the true sense of the word. But this  is not  much  in accord with the  expectations of string theory; for instance with the flatness of Calabi-Yau manifolds which are of crucial phenomenological importance. On the other hand if another perfect fluid term, now not  a curvature term for ${\mathcal{M_{E}}}$, again with $\omega=-1$ and yet with another $\nu$ accompanies our $\rho$ the observation is invariant: one pressure coefficient along extra dimensions has to be less than -2.  After \cite{eksi2ben} appeared Greene and Levin \cite{yavuzgl} reemphasized  the need for $\nu=-2$ for constant radion and they argued that  dynamical stabilization can be achieved via Casimir effect along extra dimensions, including massive contributions. At any rate here the impact of our phenomenological approach is clear; the constraint on $\nu$ remain the same and as a side nuisance  we have the above condition on $a$. 

That $\nu=-2$ in the analysis of this chapter, and $\nu\leq -2$ of  \cite{eksi2ben} for that matter, are all in close relation and accord with the recently proved  no-go theorems \cite{theo1}-\cite{theo3}. A general result of the mentioned theorems is that for a constant radion (and dilaton with a slight modification of the arguments) and observed acceleration one has to allow for violations of null energy condition and that this violation has to be generally along extra dimensions. And that it has to be {\em strong}. 

So one has to come up with objects in string theory satisfying the amendments on $\nu$ {\em and} $a$. We would like to contrast this to the {\em ease} with which a simple cosmological constant is able to accommodate the observational requirements of four-dimensional cosmology. 

We would also like to point out as a side remark that $\nu=-2$ along with $a=-2$ would mean that the lagrangian in (\ref{eq1}) along with the particular choice in (\ref{eqeben}) is invariant under both S and T duality transformations for $d=10$ and $m=3$; an observation that follows from the application of the general findings of \cite{sgc27} to the particulars of this work. 

\section{A note on dilaton coupling to conserved energy-momentum tensors}

At this point we would like to emphasize a detail about the dilaton coupling to sources that yield a conserved energy momentum tensor. For concreteness
let us assume $\mathcal{L}$ in (\ref{eq1}) is not specified. Using the equations of motion arising from varying (\ref{eq1}) with respect to the metric and $\phi$ and the fact that the energy-momentum tensor originating from $\sqrt{-q}\mathcal{L}$ is conserved one can arrive at the following  invoking the contracted Bianchi identity, as was done in \cite{sgc22}, 

\be{\label{cons1}}
\tau e^{a\phi}\left(\mathcal{L}\delta^{\mu}_{\nu}-2T^{\mu}_{\nu}\right)\nabla^{\nu}\phi=0\;.
\ee

One can read the implications  of this equation in various ways\footnote{ {One must however keep in mind that this results from the Bianchi identities and hence must be automatically satisfied via the equations of motion of the fields in $\mathcal{L}$ if we knew its form. Nevertheless the implications of (\ref{cons1}) are rather illuminating.}}. If one has $\tau=0$, this is a quite general resolution. This implies $a=2$ meaning that the fields in $\mathcal{L}$ couple minimally to the dilatonic-gravitational part of the lagrangian in (\ref{eq1}); the principle of equivalence is obeyed. We would like to point out however the following fact; let us assume $\mathcal{L}$ represents everything else, then if the dilaton couples to all of them with the same $a$ one can argue that there is still a zest of equivalence principle at work in view of this universality, albeit this would not be the usual one; it would be one generalized with the presence of the dilaton\footnote{If one insists on the usual equivalence principle one must take $a=2$. But for instance, D-branes are known to have $a=1$.}.  Nevertheless $a\neq 2$ has rather non-trivial consequences even in the absence of knowledge about the exact form of the lagrangian. So let us assume it for the sake of argument. Then, (\ref{cons1}) represents an interesting constraint on $\nabla^{\nu}\phi$ in that if it is not identically zero it has to be a vector in the null-space of the matrix 

\be
{\rm M}_{\mu\nu}\equiv \left(\mathcal{L}g_{\mu\nu}-2T_{\mu\nu}\right)=-2\frac{\partial{\mathcal{L}}}{\partial g^{\mu\nu}}
\ee

One can also approach (\ref{cons1}) in another way. One could fix the dependence of $\phi$ on the metric co-ordinates and digress on the form of $\mathcal{L}$. As pointed out in \cite{sgc22} if the dilaton depends only on time and $T_{\mu\nu}$ is diagonal we must have $\mathcal{L}=2T^{0}_{0}=-2\rho$. Elaborating on this observation still assuming that the energy-momentum tensor is of perfect fluid form we can see that if the dilaton depends on any other co-ordinate along with time, the respective pressure coefficient has to be $-1$. For instance if the dilaton is to depend on time and on the co-ordinates of extra-dimensions\footnote{ {This is still commensurable with a cosmology which is isotropic and homogeneous along the observed dimensions.}}, it must have pressure $-1$ along them, as long as $a\neq 2$. Furthermore, again in a perfect fluid approach, if the dilaton is to have non-trivial dependence on all co-ordinates we end up with $\omega=-1$ and $\nu=-1$, compatible with a d-dimensional cosmological constant or a pure dilaton potential. It is tempting to speculate that these observations are somehow related to the previously mentioned no-go theorems presented in \cite{theo1}-\cite{theo3}.

\section{A Generalization}

The analysis of section \ref{secII} does not yield a true stabilization of the radion and the dilaton, it simply studies the constraints on the parameters to have a constant dilaton and radion of unspecified value. In fact if one performs a linear stability analysis around solutions  one will find that the perturbations of both radion and  dilaton have zero mass. This is quite expected as a consequence of the fact that our toy model so far depends on  only one $\rho$;  if the righthandside's of (\ref{cieq}) and (\ref{phieq}) are to vanish any further derivative of these terms with respect to $C$ or $\phi$ also vanish at the solution. A massless excitation is not phenomenologically favoured thus we need to have true stabilization with positive masses for the radion and dilaton perturbations.

 As a general rule of thumb we need to have at least two sources to have dynamical stabilization. Both of these sources must have the same\footnote{ {Otherwise one will redshift faster than the other and we will eventually end up with a single source.}} $\omega$, which should be $-1$ to yield $B(t)=Ht+B_{o}$. So for example we can pick one of them to be like (\ref{eqeben}) and another to be a pure dilaton potential.  Let us therefore assume (\ref{eq1}) still applies but now with 

\be
e^{a\phi}\mathcal{L}=-2 e^{a\phi}\rho- 2V(\phi)\;.
\ee

This generalization will yield the following equations of motion

\begin{subequations}
\bea
\ddot{B}+\dot{k}\dot{B}&=&-e^{a\phi}\left(1+\tau\right)\rho\;-\frac{1}{2}V'\;,\label{bieq2}\\
\ddot{C}+\dot{k}\dot{C}&=&e^{a\phi}\left(\nu-\tau\right)\rho\;-\frac{1}{2}V'\;,\label{cieq2}\\
\ddot{\phi}+\dot{k}\dot{\phi}&=&\frac{1}{2}e^{a\phi}\left[T-(d-2)\tau\right]\rho\;-V-\frac{d-2}{4}V',\label{phieq2}\\
\dot{k}^{2}&=&m\dot{B}^{2}+p\dot{C}^{2} + 2 e^{a\phi}\rho + 2 V\;,\label{refk12} 
\eea
\end{subequations}

The conditions for the existence of an extremum and that $B=Ht+B_{o}$ with $H>0$ are simply given as

\begin{subequations}
\bea
0&<&-(1+\tau)U_{o}-\frac{1}{2}V'_{o}\label{sol1}\\
0&=&(\nu-\tau)U_{o}-\frac{1}{2}V'_{o}\label{sol2}\\
0&=&\frac{1}{2}\left[T-(d-2)\tau\right]U_{o}-V_{o}-\frac{d-2}{4}V'_{o}\label{sol3}
\eea
\end{subequations}

\noindent where the subscripts $o$ refer to the values at the extrema. We have also defined $U_{o}=\rho_{i}e^{a\phi_{o}-(1+\nu)pC_{o}}$ to have compact expressions. A linear stability analysis around this solution will yield the following

\be
\delta{\ddot{X}}=-F\delta{\dot{X}}-\Sigma \delta{X}
\ee

\noindent with $\delta{X}^{T}=\left(\delta B,\delta C,\delta\phi\right)$. Also $F$ is the friction matrix and has the following form

\begin{center}
\be
F=\left(\begin{tabular}{lll}
$2mH\;$ & $pH$ & $-2H$ \\ 
$\;\;0$  & $mH\;\;$ & $\;\;0$ \\ 
$\;\;0$ & $\;0$   & $mH$
\end{tabular}
\right)\ee
\end{center}

\noindent which clearly enforces damping on all equations. On the other hand the matrix $\Sigma$ responsible for the frequencies of oscillations around the extrema has the form

\begin{center}
\be{\label{eqsigma}}
\Sigma=\left(\begin{tabular}{lll}
$\;0\;$ & $\Sigma_{BC}\;$ & $\Sigma_{B\phi}$ \\ 
$\;0\;$  & $\Sigma_{CC}\;$ & $\Sigma_{C\phi}$ \\ 
$\;0\;$ & $\Sigma_{\phi C}\;$   & $\Sigma_{\phi\phi}$
\end{tabular}
\right)\ee
\end{center}

\noindent where we have

\begin{subequations}
\bea
\Sigma_{BC}&=&-(1+\nu)(1+\tau)pU_{o}\;,\\
\Sigma_{B\phi}&=&\frac{1}{2}V''_{o}+a(1+\tau)U_{o}\;,\\
\Sigma_{CC}&=&(1+\nu)(\nu-\tau)pU_{o}\;,\\
\Sigma_{C\phi}&=&\frac{1}{2}V''_{o}-a(\nu-\tau)U_{o}\;,\\
\Sigma_{\phi C}&=&\frac{1}{2}(1+\nu)\left[T-(d-2)\tau\right]pU_{o}\;,\\
\Sigma_{\phi\phi}&=&-\frac{1}{2}a\left[T-(d-2)\tau\right]U_{o}+V'_{0}+\frac{d-2}{4}V''_{o}\;.
\eea

\end{subequations}

The existence of a vanishing column in (\ref{eqsigma}) is simply a consequence of the fact that $\omega=-1$; the derivatives of the righthandsides of all (\ref{bieq2})-(\ref{phieq2}) with respect $B$ identically vanishes. In other to have stabilization we would require positive eigenvalues for $\Sigma$. But this is not the whole issue; as it stands $\Sigma$ also describes a general mixing between the perturbations which can be detrimental since it implies mixing of $\delta C$ and of $\delta\phi$ to $\delta B$. A simple and somewhat elegant way of getting around this obstacle is to assume $\Sigma_{\phi C}=0$. Requiring positive eigenvalues along with this simplifying restriction will yield

\begin{subequations}
\bea
0&<&(1+\nu)(\nu-\tau)\label{sol4}\\
0&<&-\frac{1}{2}a(T-(d-2)\tau)U_{0}+V'_{o}+\frac{d-2}{4}V''_{o}\label{sol5}\\
0&=&(1+\nu)(T-(d-2)\tau)\label{sol6}
\eea
\end{subequations}

Analysing (\ref{sol6}) we immediately see that for these equations to be consistent one has as before $T-(d-2)\tau=0$ for the other solution is $\nu=-1$ ans this is in conflict with (\ref{sol1}) used along with (\ref{sol2}). In fact from these considerations we have $\nu<-1$. Therefore we again see that for $d=10$  the $\rho$ contribution to the lagrangian is S dual\footnote{ {In this case, as opposed to the example of section II, the $\rho$ term is not T duality invariant. It better not be  because $V'$ term also contributes to the righthandside of the $C$ equations.}}. Having picked $\Sigma_{\phi C}=0$ we have achieved the following: the equations for $\delta\phi$ completely decoupled and its solutions are simply damped oscillations since $F_{\phi\phi}>0$ and $\Sigma_{\phi\phi}>0$. As a result of this one can take this solution and paste in into $\delta C$ equations where it will act as a source term: a source which asymptotically vanishes as a result of the damping. The solution of $\delta C$ thus obtained can be used in the $\delta B$ equation again as a source. Consequently the evils of mixing are somewhat circumvented. One could along with $\Sigma_{\phi C}$ also take $\Sigma_{C\phi}=0$ which completely decouples the radion and the dilaton but this is not necessary for this simple example. However We would like to stress again the fact that taking $\Sigma_{\phi C}=0$ is synonymous with the S duality of the $\rho$ term, at least for ten dimensions. Using the above intermediate result along with (\ref{sol4}) we arrive at the following

\begin{subequations} \label{hemhem2}
\bea\label{eqson}
\nu &<& -\frac{m+1}{m-1}\\
a&=&\frac{p-2+p\nu}{4}
\eea
\end{subequations}

\noindent For say $m=3$ and $d=10$ it is clear that $\nu<-2$ and $a<-2$. We have thus established that true stabilization can be achieved and that the stringent constraints on $\nu$ and $a$ remain as upper bounds. There are further consistency conditions on $V$ given as

\begin{subequations}
\bea
V'_{o}&<&0\\
V_{o}&=&-\frac{d-2}{4}V'_{o}\\
V''_{o}&>&-\frac{4}{d-2}V'_{o}
\eea
\end{subequations}

\noindent These constraints on $V$ are not too illuminating and can possibly be satisfied somewhat easily for  a wide range of models. 

\subsection{Digression on further generalizations}

In this chapter we have presented this simple generalization as an example evidencing that the constraints of section II are somewhat robust. There can be aesthetical objections to our approach since the extremum condition on (\ref{cieq2})  in essence assumes a fine tuning between the $\rho$ and $V$ terms; two contributions that possibly have nothing to do with each other. 

But again, our emphasis was not on the precise way dilaton and radion stabilization along with accelerating observed dimensions is achieved, it was on making the case for the  necessity for rather exotic sources to achieve it in general. 

Nevertheless this aestethical objection will generally be present whenever we want to achieve both radion and dilaton stabilization with few sources having $\omega=-1$. Such terms aren't exactly in abundance; a pure dilaton potential, a $\rho$ of the type we have studied and a curvature term for the extra dimensional manifold are the simplest ones that come to mind. A deeper reason for the mentioned fine tuning between sources that are at face value unrelated could be the fact that these sources become overworked in that we require both radion and dilaton stabilization from them.  In reality what we truly need is the consistent presence of only one such term for the equation of the scale factor of observed space to ensure $B=Ht+B_{o}$. Thus it is an intriguing possibility to check for sources with different pressure coefficient along observed dimensions such that the responsibilities to have accelerated observed space and stabilized radion and dilaton are separated. 

As a simple counter example to this somewhat attractive possibility we would like to digress on the impact of $(m,n)$ strings as studied in \cite{sgc22.5}. In that work there are two  contributions to the energy-momentum tensor; the winding and momentum modes of strings. Both of these sources have zero pressure along observed dimensions. On the other hand they also bring about a potential term for the dilaton equations. It can be shown that in this picture both the dilaton an the radion can be stabilized. The crucial point is that these sources will force the observed dimensions to grow as pressureless ordinary matter would; in a decelerating way. So what we need is a source with $\omega=-1$ that does not contribute to  the $C$ and $\phi$ equations. The resolution is very simple; on top of the winding and momentum mode contributions of $(m,n)$ strings  add a source  which satisfies (\ref{hemhem}). This should work since we have already seen that these conditions mean that the mentioned source is invariant under T and S dualities and thus will not contribute to the right hand sides of the dilaton and radion equations but will have an effect on the $B$ equations. The situation will be described by the following
equations

\begin{subequations}
\bea
\ddot{B}+\dot{k}\dot{B}&=&e^{-mB}S_{B}(\phi,C)-\left(1+\tau\right)e^{a\phi}\rho\;\\
\ddot{C}+\dot{k}\dot{C}&=&e^{-mB}S_{C}(\phi,C)\\
\ddot{\phi}+\dot{k}\dot{\phi}&=&e^{-mB}S_{\phi}(\phi,C)
\eea
\end{subequations}

\noindent The first term in the $B$ equations is becoming less and less relevant since the $\rho$ term does not depend on $B$. So in time we will have $B=Ht+B_{o}$. The terms $S_{C}$ and $S_{\phi}$ represent contributions from the winding and momentum modes of $(m,n)$ strings and for their explicit expressions we refer the reader to \cite{sgc22.5}. Consequently we will have dynamically stabilized radion and dilaton along with accelerated observed dimensions. However the price we pay is that the masses  of the excitations around the minima are becoming exponentially small in time even though they are always positive since the right hand sides of both $C$ and $\phi$ equations above are multiplied by $e^{-mB}=e^{-mHt}$. Furthermore this multiplicative factor can not change the location of the minima in $(C,\phi)$ space. Now since $H$ is rather small one can possibly argue in favour of such a scenario at least for now but as time evolves we would have less and less massive excitations and this is undoubtedly a source for stringent constraints on such an approach.

We can thus conclude that we typically need sources with the same $\omega$ achieving both dilaton and radion stabilization and that the aesthetical objection raised at the beginning of this subsection is, even though still standing, a bit too restrictive.

Another way to achieve stabilization without a fine-tuning between radion dependent objects and a pure dilaton potential is to include a curvature term for extra dimensions along with our $\rho$. This is possible but it will, along with (\ref{hemhem2}), require a $\mathcal{M_{E}}$ which is negatively curved.   A situation that, as we have stressed before, is not compatible with the need of string theory for Calabi-Yau manifolds.

\section{Conclusion}

We have presented stringent constraints on parameters of theories yielding accelerated observed space as well as providing stable radion and dilaton. The constraints on the dilaton couplings are just as strong as those on the pressure coefficients along extra dimensions. This can possibly be understood via the fact that the dilaton can be seen as the scale factor of a compactified eleventh dimension. The observations we have presented are related and in accord with the rather strong no-go theorems on a marriage between dark energy and extra dimensional models \cite{theo1}-\cite{theo3} in that the sources are shown to very strongly violate the null energy condition along extra dimensions. In fact the theorems mentioned not only require strong NEC violation along extra dimensions but also  that this violation has to be time dependent to allow for the observed cosmological history of the universe. Such time dependence is still a possibility within string/brane gas cosmology nevertheless we have only worked in the regime where a pure de Sitter phase has already settled in the past.

Perhaps a stronger result of these theorems is that they require non-trivial distribution of density and pressure along extra dimensions. Since we have worked with homogeneous quantities it seems the simple approach we have presented here violates this fact. Nevertheless in view of this the constraints we have presented could be seen as averages over extra dimensions and still operational. This means for instance that the average of the pressure along extra dimensions should be $-2$ and thus there must be regions where it is considerably less if the pressure -and hence the energy density- is inhomogeneously distributed. 

On the other hand, dark energy is not the only source to put stringent constraints on multidimensional theories. Recently Eingorn and Zhuk have shown that
if one assumes point like sources, toroidal extra dimensions are incompatible with classical tests  such as the perihelion advance of Mercury and gravitational frequency shift \cite{zhuk1}-\cite{zhuk3}. Even though there the sources can not provide NEC violation they are inhomogeneously distributed along extra dimensions. So it is tempting to speculate on an interplay between the theorems in \cite{theo1}-\cite{theo3} as providing an avenue for even stronger constraints on extra dimensional theories.

To conclude we reemphasize the need for a source satisfying (\ref{hemhem}) to provide for stable radion and dilaton and allowing de Sitter type expansion for observed dimensions. As we have stated, in the absence of the dilaton one can argue that these constraints can be accommodated by Casimir effect along extra dimensions \cite{yavuzgl} but it is not clear how this can be extended to dilaton gravity. Nevertheless recent research \cite{gaugino} shows that supersymmetry breaking via gaugino condensation in string gas cosmology can be the responsible actor for dilaton stabilization. It is tempting to expect that this approach, since it introduces a dilaton potential, could provide a resolution.


\begin{thebibliography}{100}
\bibitem{carroll1} Sean M. Carroll, "Why the Universe is Accelerating?", ECONF C0307282:TTH09,2003; AIP Conf.Proc.743:16-32,2005, arXiv:astro-ph/0310342.
\bibitem{BV}  R.~H.~Brandenberger and C.~Vafa, "Superstrings In The Early Universe", Nucl.\ Phys.\ B {\bf 316}, 391 (1989).
\bibitem{BV0} J.~Kripfganz and H.~Perlt, "Cosmological Impact Of Winding Strings", Class.\ Quant.\ Grav.\  {\bf 5}, 453 (1988).
\bibitem{rev1} R.~H.~Brandenberger, "String Gas Cosmology", arXiv:0808.0746.
\bibitem{rev2} T.~Battefeld and S.~Watson, "String Gas Cosmology", Rev.\ Mod.\ Phys.\  {\bf 78}, 435 (2006), arXiv:hep-th/0510022.
\bibitem{dark1} S.~S.~Gubser and P.~J.~E.~Peebles, Phys. Rev. D {\bf 70}, 123510 (2004), arXiv:hep-th/0402225.
\bibitem{dark2} S.~S.~Gubser and P.~J.~E.~Peebles, Phys. Rev. D {\bf 70}, 123511 (2004), arXiv:hep-th/0407097.
\bibitem{dark3} M.~Sano and H.~Suzuki, "Wrapped Brane Gas as a Candidate for Dark Matter", Phys.\ Rev.\ D {\bf 81}, 024042 (2010), arXiv:0907.2495.
\bibitem{rev3} R.~H.~Brandenberger, "Cosmology of the Very Early Universe", arXiv:1003.1745.
\bibitem{scale} A.~Nayeri, R.~H.~Brandenberger and C.~Vafa, "Producing a Scale-invariant Spectrum of Perturbations in a Hagedorn Phase of String Cosmology", Phys. Rev. Lett. {\bf 97}, 021302 (2006), arXiv:hep-th/0511140.
\bibitem{sgc1} G.~B.~Cleaver and  P.~J.~Rosenthal, "String Cosmology and the Dimension of Space-time", Nucl.\ Phys.\ B {\bf 457}, 621 (1995), arXiv:hep-th/9402088.
\bibitem{sgc2} M.~Sakellariadou, "Numerical Experiments in String Cosmology", Nucl.\ Phys.\ B {\bf 468}, 319 (1996), arXiv:hep-th/9511075.
\bibitem{sgc3} D.~A.~Easson, "Brane Gases on K3 and Calabi-Yau Manifolds", Int.\ J.\ Mod.\ Phys.\ A {\bf 18}, 4295 (2003), arXiv:hep-th/0110225.
\bibitem{sgc4} S.~Watson and R.~H.~Brandenberger, "Isotropization in Brane Gas Cosmology", Phys.\ Rev.\ D {\bf 67}, 043510 (2003), arXiv:hep-th/0207168.
\bibitem{sgc5} T.~Boehm and R.~Brandenberger, "On T-duality in Brane Gas Cosmology", JCAP {\bf 0306}, 008 (2003), arXiv:hep-th/0208188.
\bibitem{sgc6} R.~Easther, B.~R.~Greene, M.~G.~Jackson and D.~Kabat, "Brane Gas Cosmology in M-theory: Late Time Behavior", Phys.\ Rev.\ D {\bf 67}, 123501 (2003), arXiv:hep-th/0211124.
\bibitem{sgc7} S. Alexander, "Brane Gas Cosmology, M-theory and Little String Theory", JHEP {\bf 0310}, 013 (2003), arXiv:hep-th/0212151.
\bibitem{sgc8} B.~A.~Bassett, M.~Borunda, M.~Serone and S.~Tsujikawa, "Aspects of String-gas Cosmology at Finite Temperature", Phys.\ Rev.\ D {\bf 67}, 123506 (2003), arXiv:hep-th/0301180.
\bibitem{sgc9} A.~Campos, "Late-time Dynamics of Brane Gas Cosmology", Phys.\ Rev.\ D {\bf 68}, 104017 (2003), arXiv:hep-th/0304216.
\bibitem{sgc10} R.~Brandenberger, D.~A.~Easson and A.~Mazumdar, "Inflation and Brane Gases", Phys.\ Rev.\  D {\bf 69}, 083502 (2004), arXiv:hep-th/0307043.
\bibitem{sgc11} R.~Easther, B.~R.~Greene, M.~G.~Jackson and D.~Kabat, "Brane Gases in the Early Universe: Thermodynamics and Cosmology", JCAP {\bf 0401}, 006 (2004), arXiv:hep-th/0307233.
\bibitem{sgc12} T.~Biswas, "Cosmology with Branes Wrapping Curved Internal Manifolds", JHEP {\bf 0402}, 039 (2004), arXiv:hep-th/0311076.
\bibitem{sgc13} A.~Campos, "Late Cosmology of Brane Gases with a Two-form Field", Phys.\ Lett.\  B {\bf 586}, 133 (2004), arXiv:hep-th/0311144.
\bibitem{sgc14} S.~Watson and R.~Brandenberger, "Linear Perturbations in Brane Gas Cosmology", JHEP {\bf 0403}, 045 (2004), arXiv:hep-th/0312097.
\bibitem{sgc15} S.~Watson, UV Perturbations in Brane Gas Cosmology, Phys.\ Rev.\ D {\bf 70}, 023516 (2004), arXiv:hep-th/0402015.
\bibitem{sgc16} T.~Battefeld and S.~Watson, "Effective Field Theory Approach to String Gas Cosmology", JCAP {\bf 0406}, 001 (2004), arXiv:hep-th/0403075.
\bibitem{sgc17} F.~Ferrer and S.~Rasanen, "Dark Energy and Decompactification in String Gas Cosmology", JHEP {\bf 0602}, 016 (2006), arXiv:hep-th/0509225.
\bibitem{sgc18} M.~Borunda and L.~Boubekeur, "The Effect of alpha' Corrections in String Gas Cosmology", JCAP {\bf 0610}, 002 (2006), arXiv:hep-th/0604085.
\bibitem{sgc19} A.~Kaya and T.~Rador, "Wrapped Branes and Compact Extra Dimensions in Cosmology", Phys.\ Lett.\ B {\bf 565}, 19 (2003), arXiv:hep-th/0301031.
\bibitem{sgc20} A.~Kaya, "On winding Branes and Cosmological Evolution of Extra Dimensions in  String Theory", Class.\ Quant.\ Grav.\  {\bf 20}, 4533 (2003), arXiv:hep-th/0302118.
\bibitem{sgc21} A.~Kaya, "Volume Stabilization and Acceleration in Brane Gas Cosmology", JCAP {\bf 0408}, 014 (2004), arXiv:hep-th/0405099.
\bibitem{sgc22} S.~Arapo\~{g}lu and A.~Kaya, "D-brane Gases and Stabilization of Extra Dimensions in Dilaton Gravity", Phys.\ Lett.\ B {\bf 603}, 107 (2004), arXiv:hep-th/0409094.
\bibitem{sgc22.5} S.~Arapo\~{g}lu, A.~Karak\c{c}\i\  and A.~Kaya, "S-duality in String Gas Cosmology",  Phys.\ Lett.\ B{\bf 645}, 255 (2007), arXiv:hep-th/0611193.
\bibitem{sgc23} T.~Rador, "Intersection Democracy for Winding Wranes and Stabilization of Extra Dimensions", Phys.\ Lett.\ B {\bf 621}, 176 (2005), arXiv:hep-th/0501249.
\bibitem{sgc24} T. Rador, "Vibrating Winding Branes, wrapping Democracy and Stabilization of Extra Dimensions in Dilaton Gravity", JHEP {\bf 0506}, 001 (2005), arXiv:hep-th/0502039.
\bibitem{sgc25} T. Rador, "Stabilization of Extra Dimensions and the Dimensionality of the  Observed Space", Eur.\ Phys.\ J.\  C {\bf 49}, 1083 (2007), arXiv:hep-th/0504047.
\bibitem{sgc26} A.~Kaya, "Brane Gases and Stabilization of Shape Moduli with Momentum and Winding Stress"  Phys.\ Rev.\ D {\bf 72}, 066006 (2005), arXiv:hep-th/0504208.
\bibitem{sgc27} T.~Rador, "T and S dualities and The Cosmological Evolution of the Dilaton and the Scale Factors", Eur.\ Phys.\ J. C{\bf} 52, 683 (2007), arXiv:hep-th/0701029.
\bibitem{sgc28} A.~J.~Berndsen and J.~M.~Cline, "Dilaton Stabilization in Brane Gas Cosmology", Int.\ J.\ Mod.\ Phys.\ A {\bf 19}, 5311 (2004), arXiv:hep-th/0408185.
\bibitem{sgc29} D.~A.~Easson and M.~Trodden, "Moduli Stabilization and Inflation Using Wrapped Branes", Phys.\ Rev.\ D {\bf 72}, 026002 (2005), arXiv:hep-th/0505098.
\bibitem{sgc30} A.~Berndsen, T.~Biswas and J.~M.~Cline, "Moduli Stabilization in Brane Gas Cosmology with Superpotentials", JCAP {\bf 0508}, 012 (2005), arXiv:hep-th/0505151.
\bibitem{sgc31} S.~Kanno and J.~Soda, "Moduli Stabilization in String Gas Compactification", Phys.\ Rev.\ D {\bf 72}, 104023 (2005), arXiv:hep-th/0509074.
\bibitem{sgc32} S.~Cremonini and S.~Watson, "Dilaton Dynamics from Production of Tensionless Membranes", Phys.\ Rev.\ D {\bf 73}, 086007 (2006), arXiv:hep-th/0601082.
\bibitem{sgc33} A.~Chatrabhuti, "Target Space Duality and Moduli Stabilization in String Gas Cosmology", Int.\ J.\ Mod.\ Phys.\  A {\bf 22}, 165 (2007), arXiv:hep-th/0602031.
\bibitem{sgc34} J.~Y.~Kim, "Stabilization of the Extra Dimensions in Brane Gas Cosmology with Bulk Flux", Phys.\ Lett.\  B {\bf 652}, 43 (2007), arXiv:hep-th/0608131.
\bibitem{sgc35} M.~Sano and H.~Suzuki, "Moduli Fixing and T-duality in Type II brane Gas Models", Phys.\ Rev.\  D {\bf 78}, 064045 (2008), arXiv:0804.0176.
\bibitem{sgc36} J.~Y.~Kim, "Stabilizing Radion and Dilaton with Brane Gas and Flux", arXiv:0908.4314.

\bibitem{eksi2ben} T.~Rador, "Acceleration of the Universe via f(R) Gravities and The Stability of Extra Dimensions ", Phys. Rev. D {\bf 75}, 064033 (2007),  arXiv:hep-th/0701267.
\bibitem{yavuzgl} B.~R.~Greene and J.~Levin, "Dark Energy and Stabilization of Extra Dimensions", JHEP {\bf 0711}, 096 (2007), arXiv:0707.1062.
\bibitem{gaugino} S.~Mishra, W.~Xue, R.~H.~ Brandenberger and U.~Yajnik, "Supersymmetry Breaking and Dilaton Stabilization in String Gas Cosmology", arXiv:1103.1389.
\bibitem{theo1} D.~H.~Wesleey, "Oxidised Cosmic Acceleration", JCAP {\bf 0901}, 41 (2009), arXiv:0802.3214.
\bibitem{theo2} P.~J.~Steinhardt, D.~H.~Wesley, "Dark Energy, Inflation and Extra Dimensions", Phys. Rev. D {\bf{79}}, 104026 (2009), arXiv:0811.1614.
\bibitem{theo3} P.~J.~Steinhardt, D.~H.~Wesley, "Exploring Extra Dimensions Through Observational Tests of Dark Energy and Varying Newton's Constant", arXiv:1003.2815.
\bibitem{zhuk1} M.~Eingorn and A.~Zhuk, "Multidimensional Gravity in Non-relativistic Limit", arXiv:0907.5371.
\bibitem{zhuk2} M.~Eingorn and A.~Zhuk, "Problematic Aspect of Extra Dimensions", arXiv:0912.2698.
\bibitem{zhuk3} M.~Eingorn and A.~Zhuk, "Classical Tests of Multidimensional Gravity: Negative Result", arXiv:1003.5690.
\end{thebibliography}
\end{document}